\documentclass[fleqn,12pt,twoside]{article}
\usepackage{espcrc1,epsfig}

\usepackage{graphicx}
\usepackage[figuresright]{rotating}

\newcommand{\AmS}{{\protect\the\textfont2
  A\kern-.1667em\lower.5ex\hbox{M}\kern-.125emS}}
\newcommand{\ms}{M$_{\odot}$}
\newcommand{\zs}{Z$_{\odot}$}
\title{The early phases of Milky Way's chemical evolution }

\author{N. Prantzos\address[IAP]{Institut d'Astrophysique de Paris, 
        98bis Bd Arago, 75014 Paris}}
       
\begin{document}

\maketitle

\begin{abstract}
The earliest phases of the chemical evolution of our Galaxy are analysed in
the light of the recent VLT results (concerning abundance patterns in the
most metal-poor stars of the Galactic halo) and of stellar nucleosynthesis
calculations. It is argued that: 1) the unexpected abundance
patterns observed in Pop. II stars are not the imprints 
of an early generation of supermassive Pop. III stars; 2) among
the various suggestions made to exlain the observed abundance patterns,
nucleosynthesis in asymmetric supernova explosions appears most promising.
In the latter case, an indirect correlation between asymmetry and metallicity
is suggested by the data. Finally, the VLT data confirm two old ``puzzles'':
the existence of primary N early in Galaxy's evolution
(which constrains the mixing of protons with He-burning products
in massive stars) and the absence of 
dispersion in abundance ratios, at least up to the Fe peak,
in the early Galaxy (which bears on the timescales of homogeneisation
of the interstellar medium, but also on yield variations among massive stars).
\end{abstract}

\section{Introduction}

Galactic chemical evolution is not yet a mature dicsipline of modern
astrophysics, since many of its ingredients (large scale gas mouvements,
like inflows or outflows; interactions
between stars and gas, like large scale star formation
and supernova feedback, etc.) have no sound theoretical foundations/understanding
at present. Still, galactic chemical evolution models offer a useful
framework, inside which one may try to interpret the (everincreasing) wealth
of observational data concerning abundances in stars of the Milky Way and
other galaxies, as well as in the interstellar and even the intergalactic
medium.

In particular, studies of abundance ratios in the oldest stars of the 
Milky Way allow one to put interesting constraints on the nature of the first
stars that enriched the interstellar medium with metals. Ultimately, 
one may hope to find a distinctive imprint of the very first  
generation of massive stars (Population III), those that were born out of
primordial material containing H and He only (as well as trace amounts 
of $^7$Li, resulting from the Big Bang). The underlying assumption is that
some property of those stars (characteristic mass, rotation, explosion energy 
etc.) was quite different from the corresponding one of their more metal
rich counterparts and produced a different nucleosynthetic
pattern in their ejecta. This pattern should then be visible on the
surfaces of the next generation stars that were born from the enriched
gas and are still around today (i.e.
the oldest and most metal poor Pop. II stars).

\section{The first stars: supermassive, massive or normal?}

On theoretical grounds, it has been conjectured that Pop. III stars
should be more massive on average than stars of subsequent, metal
enriched, generations. The reason is that, in the absence of metals,
gas cannot radiate efficiently away its heat (the most efficient cooling agent
in that case being the H$_2$ molecule); thus, it cannot
reduce sufficiently its  internal pressure and collapse gravitationally.
Even in the absence of metals,
the complexity of the problem of gas collapse and fragmentation 
(involving various forms of instabilities, turbulence, magnetic fields, 
ambipolar diffusion etc.) is such that the typical mass of
the first stars is very poorly known at present
(see Glover 2004 and references therein). It may be as ``low'' as the
mass of a typical massive O star today (i.e. a few tens of \ms)
or as large as several hundreds of \ms. It is even possible that, depending
on ambient density, two ranges of typical masses co-exist for Pop. III stars,
i.e. one around a few \ms, and another around a few hundreds of \ms \ (Nakamura
and Umemura 2002). Note that, even if the first formed (zero metal) 
stars were exclusively massive, they may have induced formation of low mass
stars by compressing the surrounding gas after their supernova explosions; 
depending on the efficiency of  the mixing of the supernova ejecta with
the gas, these low mass, long lived, stars may have extremely
low  metallicities  and even no metals at all (Salvaterra et al. 2004).
Thus, even if the first stars were massive, one may still expect to
find today low mass stars of zero metallicity. 

On the other hand, an intriguing result of the Wilkinson Microwave
Anisotropy Probe (WMAP) has been interpreted as potential evidence
that the caracteristic mass of the first stellar generation was larger
than today. The optical depth along the line of sight to the last
scattering surface of the Cosmic Microwave Background
($\tau_e \ = \ \sigma_T \ c \ \int n_e dt$, where $\sigma_T$, $c$ and $n_e$ 
are the Thompson cross section, the light velocity and the free electron 
density, respectively)
was found to be $\tau_e \sim$0.17 
(Kogut et al. 2003). This, rather large, value 
suggests that the Universe must have been ionized in the past;
since hydrogen recombination is quite efficient at late times
(it becomes faster than the cosmic expansion at redshifts $<$8, i.e.
in the last 10 Gyr), the WMAP result implies that the Universe
was ionized quite early in its history.

The smallest the metallicity of a star, the more compact and hotter its is,
i.e. massive stars of very low metallicities are much more efficient ionizing
agents than their more metal rich counterparts (depending on  stellar mass,
by factors of 4-20, according to Schaerer 2002). However, a ``normal'' 
population of zero metallicity stars (i.e. having a normal Initial Mass 
Function (IMF), characterised by a typical mass $<$1 \ms) cannot produce enough
ionizing photons per baryon to account
for the observed cosmological optical depth (e.g. Sokasian et al. 2003);
a population richer in massive stars seems required. Such a population,
however, not only is an efficient ionizing agent, but also produces
metals efficiently (its metal yield is much higher than the one of a normal
stellar population) and injects large amounts of kinetic energy. 
Ricotti and Ostriker (2004) find that
the ejected metals of the very first supermassive Pop. III stars 
would bring the metallicity of the
dense regions of the intergalactic medium to unacceptably high levels,
while their kinetic energy would stop further formation of supermassive
stars, before the required degree of ionization is reached. 
A solution to those problems consists in assuming
that most of those stars implode to black holes, in which case the
ionization efficiency is decoupled from metal production
(see also Tumlinson et al. 2004) or kinetic energy injection. 
Another possibility is that the ionizing
agents were not stars, but black holes accreting from the ambient medium
(Ricotti and Ostriker 2004, Yoshida et al. 2004). But black holes
are the end points of the evolution of massive stars, so the question
of the nature of the first stars remains important even in that case.

Heger and Woosley (2002) considered in some detail the ultimate fate
of zero metallicity, non-rotating stars in the mass range 10-1000 \ms
(note that mass loss is expected to be negligible at low metallicities, thus
removing one of the - many - uncertainties of the problem). They find
that: among the Very Massive Stars (VMS, $>$100 \ms), those in the
range 130-260 \ms \ undergo pair-instability SN explosions (PISN)
and produce a distinctive nucleosynthetic pattern (see Sec. 4); 
stars with M$>$260 \ms, or with 50$<$M/\ms$<$130 implode directly
to black holes, with no metal ejection; and stars with M$<$50 \ms \
explode as SN and leave either a neutron star (M$<$25 \ms) or a black
hole after fallback of the inner parts ( 25$<$M/\ms$<$50). Needless
to say that, in view of the current difficulties to understand
core-collapse SN explosions (e.g. Janka, these proceedings), the mass limits
in the last case should be considered as very uncertain; this becomes even
more true if other, potentially important, ingredients of the problem
are considered as well (e.g. rotation, see Sec. 4).

In summary: star formation theory (and simulations) suggest that
the first stars were (perhaps Very) massive; WMAP data suggest that efficient
ionizing agents existed early in the cosmic history; and stellar 
evolution/nucleosynthesis theory predicts metal yields for a wide range
of stellar masses, subject to considerable uncertainties
(due to convection, rotation, nuclear physics 
and explosion mechanism, i.e. energetics
and ``mass-cut''). Is there any possibility to check (some of) those ideas?

\begin{figure}[!t]
\begin{center}
\psfig{figure=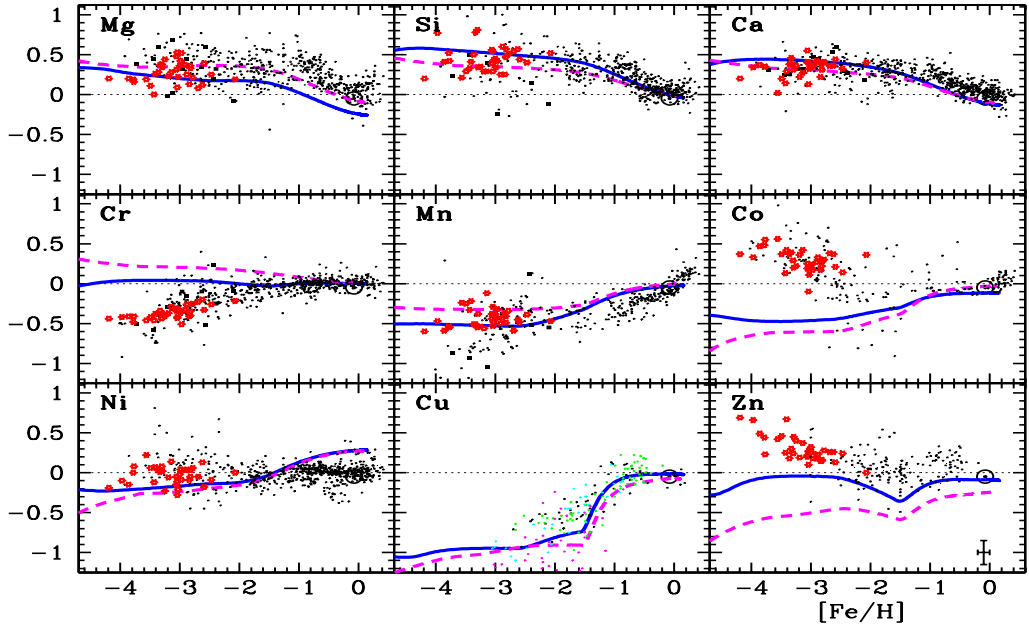,width=9cm,height=15cm,angle=-90}
\caption{Abundance ratios [X/Fe] as a function
of metallicity [Fe/H] in stars of the Milky Way; small data points
are from various sources, while the large data points at low 
metallicity are from the recent VLT survey of Cayrel et al. (2004).
Results of standard chemical evolution models, performed with two
sets of metallicity dependent massive star yields 
(Woosley and Weaver 1995, solid curves ; Chieffi and Limongi 2004, dashed
curves) are also displayed. While the behaviour of alpha elements
and Mn is correctly reproduced (at least qualitatively), 
there are large discrepancies
in the cases of Cr, Co and Zn.
}
\end{center}
\end{figure}

\section{Abundance patterns in the early Milky Way}

Around the mid-90ies, there was a rather simple picture of the expected
abundance patterns for alpha-elements (i.e. O, Mg, Si, S, Ca) and Fe-peak 
elements (i.e. Cr, Mn, Fe, Co, Ni, Cu and Zn) in the local Galaxy\footnote{
By ``abundance pattern'' we usually designate the abundance ratio of an 
element X to Fe, expressed in solar units (i.e.  
[X/Fe]=log${{({\rm X/Fe})_*}\over{({\rm X/Fe})_\odot}}$), running 
as a function of metallicity [Fe/H].}

i) Fe and Fe-peak elements are  produced by both core-collapse SN (CCSN, 
from short-lived progenitors) and thermonuclear SN
(SNIa, from long-lived progenitors), 
in a proportion of $\sim$1/3 and  $\sim$2/3, respectively, 
in the case of the solar neighborhood.

ii) alpha-elements are only produced from CCSN, thus their ratio 
$\alpha$/Fe
is expected to stay $\sim$const.$\sim$3 times solar in the old stars
of the Galactic halo (since that phase lasted for only about 1 Gyr and SNIa
progenitors had no time to evolve)
and to slowly decline down to the corresponding solar value in the more
recently formed  stars of the Galactic disk.
 
iii) the ratio of Fe-peak elements X/Fe is expected to stay constant and
equal to its solar value in stars of  all metallicities.

Observations in the mid-90ies (Mc William et al. 1995, Ryan et al. 1996)
revealed already rather 
strange patterns for some Fe-peak elements, like Cr and Co,
while Primas et al. (2000) found hints for supersolar Zn/Fe at
low metallicities. Most resently,
in the framework of ESO's Large programme ``First Stars'', 
abundances for 17 elements, from C to Zn were obtained for about 30 giant
stars with metallicities -4.1 $<$ [Fe/H] $<$ -2.7, through high resolution and
signal/noise ratio data of VLT/UVES (Cayrel et al. 2004).


The recent data confirm the previously found tendencies, concerning 
alpha elements, but also Cr, Co and Zn (Fig. 1); 
they also suggest a surprisingly small scatter in the abundance
ratios, even at the lowest metallicities (see Sec. 4).
Since the alpha-element behaviour is rather
well understood (see model curves in Fig. 1)
we focus in the following on the Fe-peak elements, leaving aside some
interesting implications of the VLT data for odd-elements, like Na and Al.

Results of nucleosynthesis of CCSN in the ``normal'' mass range
of 12 to $\sim$40 \ms \ and with progenitor metallicities from Z=0 to
Z=\zs, were published by two groups (Weaver and
Woosley 1995, Chieffi and Limongi 2004). These yields are adopted in 
a full-scale chemical evolution model of the Milky Way (originally presented
in Goswami and Prantzos 2000)
and the results are compared to the observations in Fig. 1. 
It can be seen that, while there is rather good agreement in the cases
of Mn, Ni and Cu, there is a severe problem with both sets of yields
in the cases of Cr, Co and Zn. 

At this point, let us recall how the main isotopes of the Fe-peak elements
are produced in CCSN (see WW95):

- Cr and Mn are produced through incomplete explosive Si burning, at
peak temperatures T$\sim$4-5 10$^9$ K.

- Fe is produced in complete explosive Si-burning, at
peak temperatures T$>$5 10$^9$ K. 

- Co and Ni are produced by ``a-rich freeze-out'' (i.e. at relatively
low densities) in complete explosive Si-burning.
 
- The Cu isotopes are produced in the explosion, but also by neutron
captures in the C-shell; the latter process dominates at higher metallicities
and explains the observed rise of Cu/Fe with metallicity.

- Zn is produced as radioactive Ge and Ga in ``a-rich freeze-out'', while at 
higher metallicities also by (metallicity dependent) n-captures; however,
contrary to the case of Cu, the latter effect is not observed.

In view of the above theoretical considerations, the published yields of
WW95 and CL04 and the VLT data (assuming they are not
plagued by systematic uncertainties), several questions arise:

1) Observationally, Cr/Fe and Mn/Fe behave similarly (albeit not exactly so:
they reach their  corresponding solar values at [Fe/H]$\sim$-1.8 and $\sim$0, 
respectively); theoretically, both ratios should behave similarly, still
both WW95 and CL04 overproduce Cr/Fe and predict relatively
correctly the behaviour of Mn. Why ?

2) Observations display an increasingly supersolar ratio for both Co/Fe and
Zn/Fe as metallicity decreases; on the contrary, both WW95 and CL04 
systematically underpredict those ratios at very low metallicities. Why ?

\begin{figure}[t]
\begin{center}
\psfig{figure=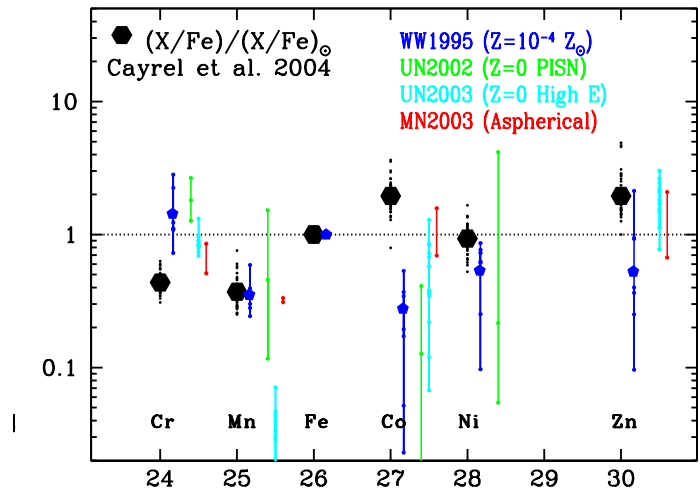,angle=-90,width=0.60\textwidth}
\caption{ Abundance ratios X/Fe (in solar units) for Fe peak elements.
Large hexagons correspond to the {\it average} values of the low metallicity
stars of Cayrel et al. (2004); values for individual stars of that survey
(giving an idea
of the dispersion) are also shown as smaller symbols on the same vertical line.
Adjacent lines connect  results for stars of different masses
calculated by various authors under different assumptions (see text); in the
case of WW95 yields, the larger symbol indicates the average over an IMF.
}
\end{center}
\end{figure}

\section{Atypical abundance patterns require atypical nucleosynthesis, but 
of what kind?}

The answers to those questions probably lie in our poor understanding of
the explosion mechanism and of the nature of the early CCSN. 
Several attempts have been made in recent years to identify key ingredients
of the stellar explosion/nucleosynthesis that would bring theory and
observations into agreement; none of those
produced convincing results up to now. In particular:

- {\it The position of the ``mass-cut''}: 
a deeper ``mass-cut'' favors products of
complete Si-burning, i.e. produces higher Co/Fe and Zn/Fe and lower 
Cr/Fe and Mn/Fe ratios; however, it also produces unacceptably 
large Ni/Fe ratios and should be 
ruled out (e.g. Nakamura et al. 1999)

-  {\it Very massive stars exploding as PISN}: they
do not produce enough Zn 
(Heger and Woosley 2002,  Umeda and Nomoto 2002), because they 
undergo very little complete explosive
Si-burning, and should be ruled out.

- {\it High energy explosions} (``hypernovae'', with kinetic energy values 
higher than
the canonical one of $\sim$10$^{51}$ ergs): they  produce large amounts of Zn,
but also too much Ni, so that a combination of mixing and fallback has to be
introduced to bring agreement with observations (Nakamura et al. 2001,
Umeda and Nomoto 2002); however, Mn/Fe and Co/Fe are always undeproduced 
in that case (i.e. obtained values are much lower than observed ones in 
extremely metal poor or EMP
stars), which should also be ruled out (see also Umeda et al. 2004,  
and in particular their Fig. 2)

- The precise value of {\it the electron mole fraction Y$_e$} 
in the nucleosynthesis
region of Fe-peak elements plays also an important role in shaping the final
abundance ratios, like the ones of mono-isotopic and odd 
Co and Mn (Nakamura et al. 2001; 
Umeda and Nomoto 2002, 2003); although a careful treatment of all the weak 
interactions improves the situation with Zn 
(avoiding at the same time excessive Ni overproduction, Frohlich et al. 2004), 
it certainly cannot help with the
other elements and is not the answer to the problem.

- {\it Low mass progenitors (11-12 \ms)}: they are more compact than
their more massive counterparts and have 
steeper density profiles; they provide excellent results for 
both Cr/Fe and Co/Fe,
but not for Zn/Fe (A. Chieffi, private communication) 

- {\it Asymmetric (bipolar) CCSN explosions of rotating stars}
produce a more satisfatory picture
(Maeda and Nomoto 2003):
material ejected along the rotation (jet) axis has high entropy
and is found to be enriched
in products of a-rich freeze-out (Zn and Co), as well as Sc (which is 
generically undeproduced in spherical models), while Cr/Fe and Mn/Fe
are found depleted. The energy of the explosion and the position of the
mass-cut also play a role in determining the absolute yield of $^{56}$Ni
and the $\alpha$-element/Fe ratio. This kind of models seems at present
the most promising, but this is not quite unexpected (since they have 
at least one more degree of freedom w.r.t. spherical models).

Some of the results of the aforementioned models are displayed in Fig. 2
and are compared with the VLT data for EMP stars.

Whatever the answer to the problem of the peculiar abundances
of EMP stars turns out to be, the next question, never explicitly stated 
but naturally  arising in the context of galactic chemical evolution
is: ``in what sense were the CCSN of the subsequent generations different
from the very fist ones?''. In other terms, why the abundance ratios
of Cr/Fe, Co/Fe and Zn/Fe became ``normal'' (i.e. solar) in later stages
of the evolution (but still before the appearance of SNIa)?

If the answer to the first problem were ``Very massive stars'' 
(of a few hundred \ms) then the second question has also a natural answer,
which is: ``different typical mass produces different nucleosynthesis''. 
However, VMS exploding as PISN are, at present, ruled out.
All the other potential answers (e.g. energy, mass-cut, 
rotation) concern {\it normal
massive stars} of a few tens of \ms \ and it is not at all clear how such
stellar properties
could be linked to metallicity. Indeed, the VLT data suggest a smooth variation
of Cr/Fe and Co/Fe with Fe/H and, taking into account the halo metallicity
distribution function (e.g. Prantzos 2003), one sees that a large fraction
of the halo stars (up to $\sim$20\%) may display such peculiar patterns. 
The conclusion is that the stellar property responsible for those peculiar 
patterns did not characterise only a single first generation of CCSN but
it varied smoothly with metallicity.
In the case of rotation, one may understand such a variation (at least
qualitatively) as resulting from loss of angular momentum due to
mass loss (e.g. Meynet et al. 2004): mass loss becomes more important 
in more metallic stars and reduces their final rotation rate, making their
nucleosynthesis less ``exotic''.

This picture is at odds with the idea 
(Ryan et al. 1996, Tsujimoto and Shigeyama 1998, 
Umeda and Nomoto 2003) that all the EMP stars were formed
near a single (or very few) CCSN of the first generation and  their 
metallicity was fixed by the amount of interstellar gas mixed with the SN
ejecta; according to that idea, the more energetic CCSN would mix with more 
hydrogen and produce lower
Fe/H ratios than the ones with smaller energy. However, it is hard to
understand in that framework the observed variation of the abundance of an
odd element like Na with metallicity in EMP stars (Cayrel et al. 2004); 
such variation is much
easier to interperet in the conventional framework of a well-mixed ISM
with metallicity
dependent yields for odd elements.

Those questions are naturally related to the issue of dispersion in
the abundance ratios of EMP stars. Simple-minded arguments suggest that
dispersion should increase at low metallicities, under the 
explicit assumption
that  low metallicities correspond to such early times that complete 
mixing of SN ejecta with the interstellar medium is impossible
(Argast et al. 2002, Karlsson and Gustafsson 2001). In order to 
actually see such a dispersion, 
variations in abundance ratios among SN of different masses should be
sufficiently large (say, larger than typical observational statistical 
uncertainties of 0.1-0.2 dex in the VLT data of Cayrel et al. 2004).

The VLT data reveal very small scatter in the abundance ratios of EMP stars,
compatible with observational uncertainties and in full agreement
with the results of a previous study (Carretta et al. 2002). This could
mean that i) mixing timescales are (much) shorter than typical chemical
evolution timescales at metallicities down to [Fe/H]$\sim$-4, or ii)
variations in abundance ratios of SN of different masses and energies
are sufficiently small, or a combination of (i) and (ii).

\begin{figure*}[htb]
\begin{center}
\psfig{figure=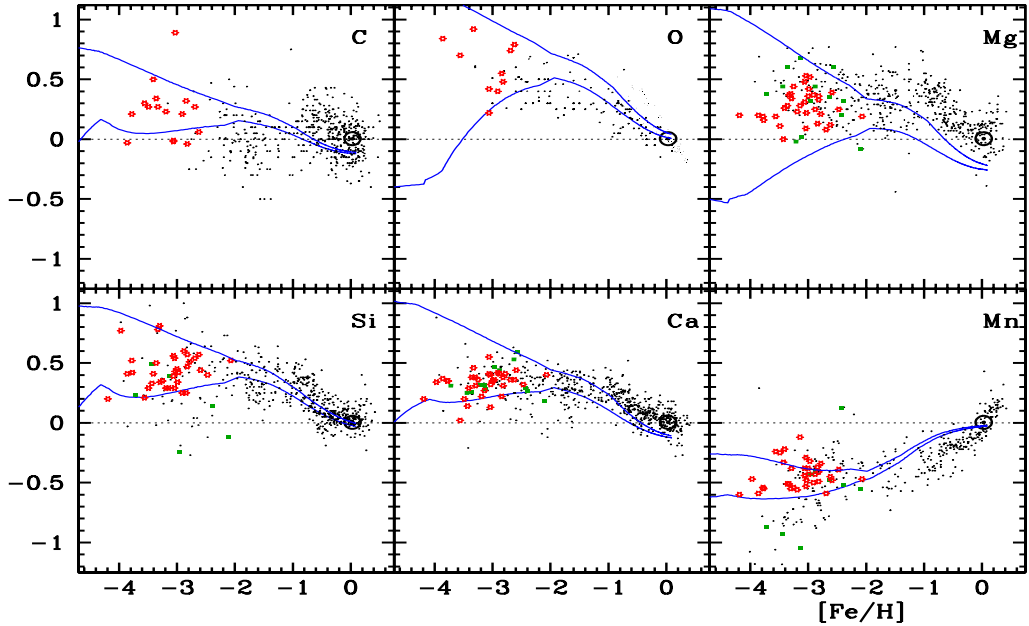,width=9cm,height=15cm,angle=-90}
\caption{Abundance ratios [X/Fe] as a function
of metallicity [Fe/H] in stars of the Milky Way; small data points
are from various sources, while the large data points at low 
metallicity are from the recent VLT survey of Spite et al. (2004) for C and O
and Cayrel et al. (2004) for the others. Curves correspond to $\pm$2 $\sigma$
dispersion calculated in a model with WW95 metallicity dependent yields
(see text); in most such models dispersion generically increases at
low metallicities, whereas observations show no such  trend.
}
\end{center}
\end{figure*}

Despite our current ignorance of the timescales of mixing
and early chemical evolution, case (i) cannot be the whole truth, since
large variations in abundance ratios are observed in the case of r-elements,
also produced from (perhaps some class of) CCSN (see Ishimaru et al.
these proceedings). Case (ii) obviously offers the opportunity to constrain
variations in yield ratios of CCSN of different masses and metallicities
(see Fig. 3 and Ishimaru et al. 2003). In that respect, it should
be noted that the ad-hoc mechanism of mixture and fallback of the
inner SN ejecta (adopted in   Nakamura et al. 2001, Umeda and Nomoto 2002 etc.)
may reduce considerably the scatter in hydrostatic/explosive or 
explosive/explosive element ratios and
lead ``naturally'' to the absence of dispersion as a function of metallicity.

\begin{figure}[t]
\begin{center}
\psfig{figure=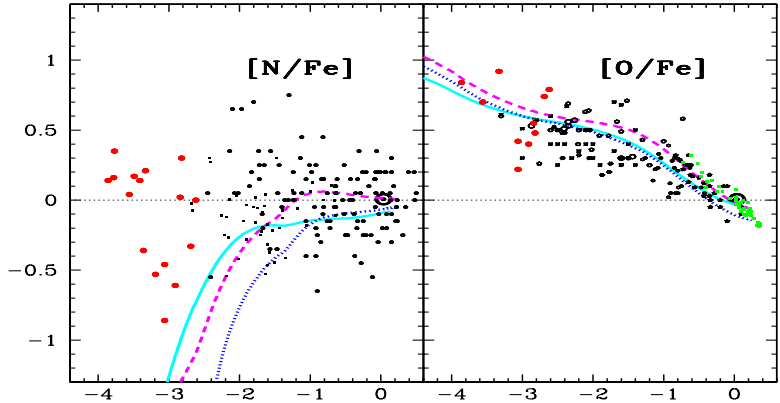,height=0.7\textwidth,width=6cm,angle=-90}
\caption{Evolution of O and N in solar neighborhood with three sets of stellar
yields: vdHG97 + WW95 ({\it solid curve}), MM02 with rotation ({\it dashed
curves}) and MM02 with no rotation ({\it dotted curves}). The first two cases,
producing primary N from intermediate mass stars
(from HBB and rotation, respectively), lead to similar results.}
\label{fig:toosmall}
\end{center}
\end{figure}

\section{``Primary'' nitrogen in the early Galaxy?}

Nitrogen is the main product of the CN cycle operating in core H-burning
in intermediate and high mass stars, in shell H-burning in stars of all masses
and in the putative Hot-Bottom Burning (HBB) in massive (4-8 M$_{\odot}$)
Asymptotic Giant Branch (AGB) stars. 
In the former two cases N is produced from the initial C and O
entering the stars, whereas in the latter from the C produced by the 3-$\alpha$
reaction in the He-shell (and brought to the envelope by convective motions
during thermal pulses). Thus, in the first two cases N is produced as a {\it
secondary} (its yield being quasi-proportional to initial stellar metallicity)
and in the latter as a {\it primary} (its yield being quasi-independent
of metallicity). On the other hand,
rotationally induced mixing may mix protons to He-burning regions 
and produce primary N in {\it rotating  stars} 
of {\it all masses} and metallicities, but mostly
in AGBs of low metallicity, as found by
Meynet and Maeder (2002, MM02). 

Due to the different evolutionary timescales between the progenitors of
CCSN and AGBs, one expects then the N/Fe ratio to decline below its
solar value early in Galaxy's evolution (at sufficiently low metallicities).
However, a VLT sample of ``unmixed''\footnote{Stars still containing
their initial Li at their surfaces, and so, presumably, also their initial
N, unaffected by any internal operation of the CNO cycle.} red giants 
shows a $\sim$const. N/Fe
ratio (i.e. primary N) down to [Fe/H]=-4 (Spite et al. 2004).

The VLT results are compared in Fig. 4 to the results of standard 
chemical evolution calculations performed with 3 sets of yields:
1) Massive stars from WW95 and AGBs with HBB from van den Hoek and
Groenewegen (1997, vdHG97); 2) MM02 yields for non-rotating stars and
3) MM02 yields for rotating stars.
The N yields of the rotating stars of MM02 (Case 3) lead to an 
evolution similar to the one obtained with the HBB yields of vdHG97 (Case 1): 
N behaves as primary with respect to Fe, but starting only at [Fe/H]$\sim$--2
(i.e. when intermediate mass AGB stars start enriching the ISM).

The recent VLT results imply that substantial primary N from massive stars 
(which dominate at the lowest  metallicities and early times) is required; 
the corresponding mechanism has not
been found up to now.  Alternatively, the timescales of simple models
of early Galaxy's evolution, which are not seriously constrained by
other observations, should have to be revised (see Prantzos 2003 
for the possibility of such a revision).

\end{document}